\documentclass[3p,times]{elsarticle}

\biboptions{sort&compress}
\usepackage{graphics}
\usepackage{graphicx,colordvi}
\usepackage{amssymb}
\usepackage{xcolor}
\def\d{\hbox{d}}

\def\sin{\hbox{sin}}

\def\sinh{\hbox{sinh}}

\def\exp{\hbox{exp}}

\def\sinh{\hbox{sinh}}

\def\siml{\hbox{\kern.1em \lower.6ex \hbox{$\sim$} \kern-1.12em
          \raise.6ex \hbox{$<$} \kern.1em }}
\def\simg{\hbox{\kern.1em \lower.6ex \hbox{$\sim$} \kern-1.12em
          \raise.6ex \hbox{$>$} \kern.1em }}

\begin{document}

\begin{frontmatter}

\title{Level density within a micro-macroscopic approach}

\author[inr]{A.G.~Magner\corref{corr1}}
\ead{magner@kinr.kiev.ua}

\author[inr]{A.I.~Sanzhur}
\author[inr]{S.N.~Fedotkin}
\author[inr]{A.I.~Levon}
\author[cyc]{S.~Shlomo}

\cortext[corr1]{Corresponding author}

\address[inr]{Institute for Nuclear Research, 03028 Kyiv, Ukraine}
\address[cyc]{Cyclotron Institute, Texas A\&M University,
College Station, Texas 77843, USA}

\begin{abstract}

  Statistical level density $\rho(E,A)$ is derived for nucleonic system with
  a given energy $E$, particle number $A$ and other integrals of motion 
  in the micro-macroscopic approximation
  beyond the standard saddle-point method of the Fermi gas model. 
  This level density 
  reaches the two limits; the well-known Fermi gas 
    grand-canonical ensemble limit 
  for a large entropy
  $S$ related to large excitation energies,
  and the finite micro-canonical limit for a small combinatorical entropy $S$
  at low excitation energies.  
    The
    inverse level
    density parameter $K$ as function of the particle 
  number $A$
  in the semiclassical 
  periodic orbit theory, taking into account 
      the extended Thomas-Fermi
  and Strutinsky shell corrections,
  is calculated
  and
  compared with experimental data. 

  \end{abstract}


\end{frontmatter}

\section{Introduction}
\label{itrod}

Many properties of heavy nuclei can be to large extent described
in terms of the statistical level density
\cite{Be36,Er60,GC65,BM67,St72,Bj74,Ig83,So90,Sh92,Ju98,AB00,AB03,EB09,Gr13,AB15,AB16,ZS16,KZ16,KS18,ZK18,ZH19,Gr19,KZ20,KS20}. 
Usually, the level density 
$\rho(E,A)$, where $E$ and $A$, 
are the energy and
nucleon number, respectively,
is calculated by
the inverse Laplace transformation of the partition function
$\mathcal{Z}(\beta,\alpha)$
of the corresponding
Lagrange multipliers.  Other integrals of motion, e.g., 
the orbital angular momentum projection $M$, or separately,
the number of neutrons and
protons instead of $A$ can be considered in a similar way by
introducing other Lagrange multipliers.
Within the grand
canonical ensemble, 
the standard saddle-point method (SPM)
was used for the integration
over all variables,
including 
$\beta$, which is related to
the total energy $E$ \cite{Er60,BM67}.
This method assumes a large
excitation energy $U$, so that the temperature $T$ can be defined
through a  well-determined
saddle point
in the integration variable
$\beta$ for a finite Fermi 
    system of large particle numbers.
However,
many experimental data are also related 
to the low-lying part of the
excitation energy $U$,
where  such a saddle point does not exist.
For presentation of experimental data on nuclear spectra, the cumulative
number of quantum levels
below a given
excitation energy $U$ 
is conveniently often used
for a statistical analysis 
\cite{Ze96,Go11,ML18}, e.g., of
the experimental data on the collective excitation energies of
rare-earth and actinide
nuclei excited in two-neutron transfer (p,t) reactions
\cite{Le19a}.
For calculations of 
this cumulative 
number one has to
integrate the level density
over a large interval of the excitation energy $U$ from small
values where
there is no thermodynamical equilibrium 
to large
values where the standard 
approach can be successfully
applied in terms of the temperature $T$ in a finite Fermi system.
Therefore, to simplify the level density, 
$\rho(E,A)$, calculations 
we 
will integrate over the Lagrange multiplier
$\beta$ in the inverse Laplace
  transformation of the partition function
 $\mathcal{Z}(\beta,\alpha)$  
  more
 accurately
  beyond the SPM. It is 
  worthwhile to unify the standard Fermi-gas model
  for large excitation energies $U$ with the empiric constant
  temperature model 
  at small $U$ as suggested in Ref.~\cite{GC65}, see also
  Refs.~\cite{ZS16,ZH19,ZK18}.

   More  general microscopic formulation of the energy level density 
 for mesoscopic systems, in particular for nuclei,  which removes
 the singularity at small excitation energies, is discussed 
in Ref.~\cite{ZH19}, see also references therein.  
One of the microscopic ways for accounting
     for interparticle interactions beyond the mean field (shell model)
     in the level density calculations
     was suggested within the Monte-Carlo Shell Model \cite{Or97,AB03,AB15}.
     Another successful 
     approach for taking into
     account the interparticle
         interactions above the simple shell model
      is given by
     the moments 
     method \cite{KZ16,Ze16,ZK18,KZ20,Ze96}. The main ideas are based on
     the random matrix theory, see Refs.~\cite{Po65,Ze96,Ze16,Me04}.

A micro-macroscopic approximation (MMA) which unifies
micro-  and  macroscopic ensembles,
for  the
statistical level density $\rho$ was 
suggested in Ref.~\cite{KM79}.
This MMA approach is based on  
the Strutinsky shell correction method \cite{St67,BD72,BK72} within
the Landau-Migdal
quasiparticle theory called as the Finite Fermi System
Theory 
\cite{MI67,HS82}.
A mean field potential is used for
calculations of the
energy shell corrections, $\delta E$. The total nuclear energy, $E$,  is the
sum
of these corrections and smooth macroscopic liquid-drop component \cite{MS69}
which can be well approximated by the Extended Thomas-Fermi ((E)TF)
approach
\cite{BG85,BB03}. Thus, 
    within the semiclassical approximation to the
    Strutinsky shell correction method,
    the interactions between particles, in particular interparticle
    collision-like
    interactions, averaged
over particle numbers, 
    i.e. over many-body microscopic quantum states 
    in realistic
nuclei, 
are approximately taken into
account through the  
extended Thomas-Fermi component beyond the mean field.
In this way, one can present \cite{KM79} 
the level density $\rho$  
in terms of the
modified Bessel function
of the entropy variable in the case of
small
 thermal excitation energy $U$ as compared to the
 rotational energy
    $E_{\rm rot}$. 
The 
shell-correction method 
\cite{St67,BD72}
was applied \cite{KM79} for studying
the shell effects in the nuclear moment of
inertia. 
For a deeper understanding of the 
correspondence between the classical and the quantum approach,
it is also worthwhile to analyze the shell effects in the level density $\rho$ 
within the semiclassical periodic-orbit theory (POT) 
\cite{BB72,SM76,SM77,BB03,MY11}.

In the present
study we 
extend, in a simple transparent way, the MMA approach \cite{KM79}
for the description of shell effects in
terms of the level density 
itself from small to large  
excitation energies $U$, see also 
    detailed derivations accounting for 
    the
    rotational and isotopically asymmetric effects 
    in Refs.~\cite{PRC,IJMPE}.
The level
density parameter $a$
is one of the key quantities
under intensive experimental and theoretical discussions
\cite{Ig83,Sh92,ZS16,KS18,EB09,KS20}.
Properties of the smooth inverse level density parameter, $K=A/a$, as 
function of the nucleon number $A$
have been studied
within the framework of the
self-consistent ETF approach \cite{Sh92,KS18}.
  However,
shell 
effects in the statistical level density is still
an attractive subject, especially important near nuclear magic numbers.

 The structure of the paper is the following.
In Sec. \ref{MMA}, the level density $\rho(E,A)$ is derived
within the MMA for the one-component nucleon system 
by using the POT. We discuss general properties of
the statistical level density
$\rho(S)$ and its approximations as function of the entropy $S$.
In Section \ref{resdisc}, we compare our analytical  MMA results for the level
density $\rho(E,A)$, and the inverse level density parameter $K$
with experimental
data for several nuclei  
as typical examples.
Our Conclusions are presented in Section \ref{concl}.
Some details of the periodic orbit theory
are presented in \ref{appA}.

\section{Micro-macroscopic approach}
\label{MMA}

 For 
a statistical
description of level density of a nucleus in
  terms of the conservation variables --
the total energy, $E$, and nucleon number,  $A$, 
one
can begin with
the microcanonical expression for the level density,
\begin{equation}\label{dendef1}
\rho(E,A)=
\sum\limits_i\!\delta(E-E_i)~\delta(A-A_i)
\equiv
\int \frac{\d \beta \d \alpha}{(2\pi i)^2}~\exp[S(\beta,\alpha)], %e^{S},
\end{equation}
where $E_i$ and $A_i$ present the system spectrum,
$S(\beta,\alpha)=\ln \mathcal{Z}(\beta,\alpha)
+\beta E -\alpha A~$
is the entropy with $\mathcal{Z}(\beta,\alpha)$ 
being the
partition function. 
Integrating over $\alpha$ for a  given $\beta$ by the
standard SPM in Eq.~(\ref{dendef1}), 
we use the expansion for this entropy
as: $S(\beta,\alpha)=S(\beta,\alpha^\ast)
+\left(\partial^2 S/\partial \alpha^2\right)^\ast
\left(\alpha-\alpha^\ast\right)^2/2+\ldots~.$
The first order term of this expansion disappears because
the Lagrange multiplier, $\alpha^\ast$,
is defined by the saddle point 
condition,
\begin{equation}\label{Seqsd}
\left(\frac{\partial S}{\partial \alpha}\right)^\ast\equiv
\left(\frac{\partial \ln \mathcal{Z}}{\partial \alpha}\right)^\ast-A=0~.
\end{equation}

Introducing, for convenience, the potential $\Omega=-\mbox{ln}\mathcal{Z}/\beta$,
one can define the system partition function $\mathcal{Z}$ through
\begin{equation}\label{Ompot}
  \Omega=\Omega_0- \frac{a}{\beta^2}~,\qquad 
      \Omega_0=E_0-\lambda A~,
\end{equation}
where $E_0$ is the energy  of a non-excited ground-state system,
$a$ is the level density parameter,
 and $\lambda=\alpha^\ast/\beta$ is associated
 with the chemical potential as function of the particle number $A$
 at arbitrary $\beta$
through Eq.~(\ref{Seqsd}).
Substituting the expansion of the entropy near the saddle point  
into Eq.~(\ref{dendef1}),
and
taking the error integral over $\alpha $ in the
infinite
limits, 
one obtains
\begin{equation}\label{rhoE1}
\rho(E,A) \approx 
\int \frac{\d \beta~\beta^{1/2}}{2\pi i~\sqrt{2\pi}}
~\mathcal{J}^{-1/2}
\exp\left(\beta U + \frac{a}{\beta}\right),
\end{equation}
where $U=E-E_0$ is the  excitation energy.   In the thermodynamic limit   
    for a large excitation
    energy $U$ 
      one finds the temperature $T=1/\beta^\ast$ through the saddle point
    $\beta=\beta^\ast$,
    and the well-known
    expression 
    for the grand-canonical  
    potential $\Omega$,  $\Omega^\ast=\Omega_0 - aT^2=\Omega_0 - U$. 
    In this limit, $\lambda=\alpha^\ast T$ is the usual chemical potential
      at the thermodynamic equilibrium with temperature $T$.
In Eq.~(\ref{rhoE1}), the one-dimensional 
 Jacobian determinant $\mathcal{J}$
($c$ number) is taken 
 at the saddle point (\ref{Seqsd}), i.e.,
\begin{equation}\label{Jac1}
\mathcal{J}\equiv\beta \left(\frac{\partial^2 S}{\partial \alpha^2}\right)^\ast
\equiv 
-\left(\frac{\partial^2\Omega}{\partial \lambda^2}\right)^\ast
=
\mathcal{J}_{1} + \mathcal{J}_{2}~,~\quad
\mathcal{J}_{1}=-\frac{\partial^2\Omega_0}{\partial\lambda^2},\quad
 \mathcal{J}_{2}=\frac{1}{\beta^2}\frac{\partial^2a}{\partial \lambda^2}~.
\end{equation}
The
  level density parameter $a$
  can be expressed in terms of the
  ETF level density
  and shell correction,
\begin{equation}\label{a}
  a=\frac{\pi^2}{6}g(\lambda),~\quad g\left(\lambda\right)=
\tilde{g}(\lambda)+
\delta g(\lambda),
\end{equation}
where $\tilde{g}(\varepsilon)$ is the Strutinsky smooth
%s.p.
level density,
approximately equal to the ETF one 
(including the
spin or
spin-isospin degeneracy). 
 The slightly averaged
    oscillating (shell) component $\delta g(\varepsilon)$
of the level density
$g\left(\lambda\right)$ 
can be calculated 
using the shell-correction method 
\cite{BD72} 
or  approximately, within the semiclassical periodic orbit theory,
see \ref{appA} and Refs.~\cite{SM76,BB03}. 
    Thus, for the level density parameter $a$
one has a decomposition of the Strutinsky
shell correction method \cite{St67,BD72}:
\begin{equation}\label{aSCM}
a=\tilde{a} + \delta a~,\quad \tilde{a}=\frac{\pi^2}{6}\tilde{g},\quad
\delta a=\frac{\pi^2}{6} \delta g~,
\end{equation}
    where $\tilde{a}$ is the smooth ETF component of the level
    density parameter $a$,
$\tilde{a}=a^{}_{\rm \tt{ETF}}$, and
$\delta a$ is its shell correction.
Both components of
    the level density $g\left(\lambda\right)$ are taken at 
the chemical potential $\varepsilon \approx \lambda$. 
The shell correction
part is averaged
 with Gaussian having the average parameter $\gamma $ which is much
 smaller than the distance between major shells.
Using the mean field  
expressions for the energy $E_0$ 
and particle number
$A$  in terms of the
level density
$g(\varepsilon)$,  one has
\begin{equation}\label{conseq}
  E_0=\int_0^\lambda \d \varepsilon\varepsilon~g(\varepsilon), \qquad
  A=\int_0^\lambda \d \varepsilon~g(\varepsilon)~.
\end{equation}
For the Jacobian component $\mathcal{J}_{1}$, one arrives
at the approximation
of smooth 
ETF level density, $g 
\approx \tilde{g}$, neglecting its derivatives
contribution
\cite{BM67},
$\mathcal{J}\cong \mathcal{J}_{1} \approx g(\lambda)$~.
 This approximation
is named below as the case (i).
According to Eq.~(\ref{a}),  for the case (ii) of the
dominating second term (\ref{Jac1})
of the Jacobian $\mathcal{J}$  one writes
\begin{equation}\label{Q}
\mathcal{J}\approx \mathcal{J}_{2}
\approx (\pi^2/6)g^{\prime\prime}(\lambda)/\beta^2\approx  (\pi^2/6)
\delta g^{\prime\prime}(\lambda)/\beta^2 ,
\end{equation}
 see Eq.~(\ref{d2Edl2}) for $\delta g^{\prime\prime}(\lambda)$.
 For following derivations it is helpful to introduce the Jacobian
 ratio\footnote{We shall present 
  the main case
  of $\delta E<0$
  near the minimum of 
   shell correction energy, as  mainly applied below.
  For the case of a
  positive $\delta E$ we change, for convenience, signs so that we will get
  $\xi>0$.} given by 
\begin{equation}
  \xi =\mathcal{J}_{2}/\mathcal{J}_{1}\approx (\pi^2/6)
  \delta g^{\prime\prime}(\lambda)/\beta^2 \tilde{g}(\lambda)
  \sim
 (4\pi^6/3)(T/\lambda)^2 A^{1/3}\mathcal{E}_{\rm sh}~,\qquad  
  \mathcal{E}_{\rm sh}=-\delta E~A/E_{\rm \tt{ETF}}~,
  \label{ratio}
  \end{equation}
where $T=1/\beta^\ast=\sqrt{U/a}$ is the temperature evaluated
through
the saddle point
$\beta=\beta^\ast$ , $\delta E\approx \delta E_{\rm scl}$
  is the semiclassical POT 
  energy shell correction,
    Eq.~(\ref{dedgnp}),
  for major shell structure \cite{SM76,SM77,BB03,MY11},
 and $E_{\rm ETF}$ is the smooth ETF energy,
  $E_{\rm ETF}
  \sim
  \tilde{g}(\lambda)\lambda^2$.
    The derivatives $\delta g^{\prime\prime}$
    are given in
    Eq.~(\ref{d2Edl2}).
  For typical values of parameters
  $\lambda=40$ MeV, $A\sim 200$, and 
     relative energy shell correction $\mathcal{E}_{sh}\sim 0.2-2.0$
  \cite{MSIS12},
  one finds the estimate 
   $\xi\sim 1 - 10$ for temperature $T\sim 1$ MeV,
    much smaller than $\lambda$,
    and $\xi\sim 0.01 - 0.1$ for very small temperature $T\sim 0.1$ MeV.
     In line with the 
  shell correction method \cite{BD72} and extended Thomas-Fermi
  approach \cite{BB03},
    these values are given finally using the realistic
  smooth energy $E_{\rm ETF}$ for which the binding energy  approximately equals 
  $E_{\rm ETF}+ \delta E$.

For calculations of the Jacobian $\mathcal{J}$, Eq.~(\ref{Jac1}),
we will consider two different limiting cases 
    (i) and (ii) (see above and more details in
 Refs.~\cite{PRC,IJMPE}).
In 
the case (i) of small contribution of the
$\beta$-dependent component $\mathcal{J}_{2}$ of the Jacobian
(\ref{Jac1}), 
$\xi \ll 1$, Eq.~(\ref{ratio}), 
that is related, for instance, to small heat excitations with
respect to the collective
(rotational) motion, we 
will approximate the Jacobian $\mathcal{J}$ by   
the constant $\mathcal{J}_{1}$ (independent of $\beta$), $\mathcal{J} \approx
\mathcal{J}_{1} $. Then, one takes the Jacobian factor
$\mathcal{J}^{-1/2}$ off the
integral in Eq.\ (\ref{rhoE1}), done as in Ref.~\cite{KM79}.
Transforming $\beta$ to a new variable $\tau=1/\beta$, we
recognize
in Eq.~(\ref{rhoE1}) the standard Laplace transform.  
 Evaluating this
 integral over $\tau$, one finally
 arrives at
  \begin{equation}\label{denbes}
    \rho \approx \rho^{}_{\tt{MMA}}(S)
    =\overline{\rho}_\nu~f_\nu(S)~,~~~f_\nu(S)=
  S^{-\nu}I_{\nu}(S)~,
  \end{equation}
  where $S$ is the  
     entropy,
  $S=2\sqrt{a~U}~$  
     with $U$ being the excitation energy defined above, $U=E-E_0$.
     This entropy $S$ 
      can be 
     expressed in terms of the level density parameter
     $a$ and excitation energy $U$ as in the mean field approach.
     However, the level density parameter  
     $a$ [Eq.~(\ref{aSCM})]
     in our MMA approach contains the ETF part $\tilde{a}$
     related to the liquid drop energy through the ETF level density
     $\tilde{g}$.
In Eq.~(\ref{denbes}),  $\overline{\rho}_\nu$ is a constant, independent of
the entropy $S$,
$~\overline{\rho}_{\nu}=
2 a^{\nu}
\pi^{1-\nu}\left\vert\mathcal{J}^{(2\nu-2)}_{1}\right\vert^{-1/2}$~, where
$2\nu-2$ is the determinant dimension
 [this superscript
is omitted in
Eqs.~(\ref{rhoE1}) and (\ref{Jac1})].
The modified Bessel function $I_\nu(S)$ of the order
of $\nu$, 
$\nu=(n+1)/2$,
is determined by the number of
integrals of motion $n$
($n=2$, and therefore, $\nu=3/2$
for this case of the
two integrals of motion  $E$ and $A$ ).
Expression (\ref{denbes}) 
is written
  in a general 
  form for arbitrary number of integrals of
  motion $n$.
  For the specific case  
  $n=2$ here
  ($\nu=3/2$), 
  within the same case (i) for relatively small shell-corrections
  contributions,
  named below also as the
    MMA1 approximation, one obtains
\begin{equation}\label{ld32}
\rho_{\tt{MMA1}}(E,A)=a\sqrt{2\pi/3}~I_{3/2}(S)/S^{3/2}~,\qquad (i);
\end{equation}
see also Ref.~\cite{KM79}
for the case of $n=3$ ($\nu=2$, $\rho=\rho(E,A,M)$) 
in Eq.~(\ref{denbes}).

\begin{figure}
  \begin{center}
    \includegraphics[width=0.6\columnwidth]{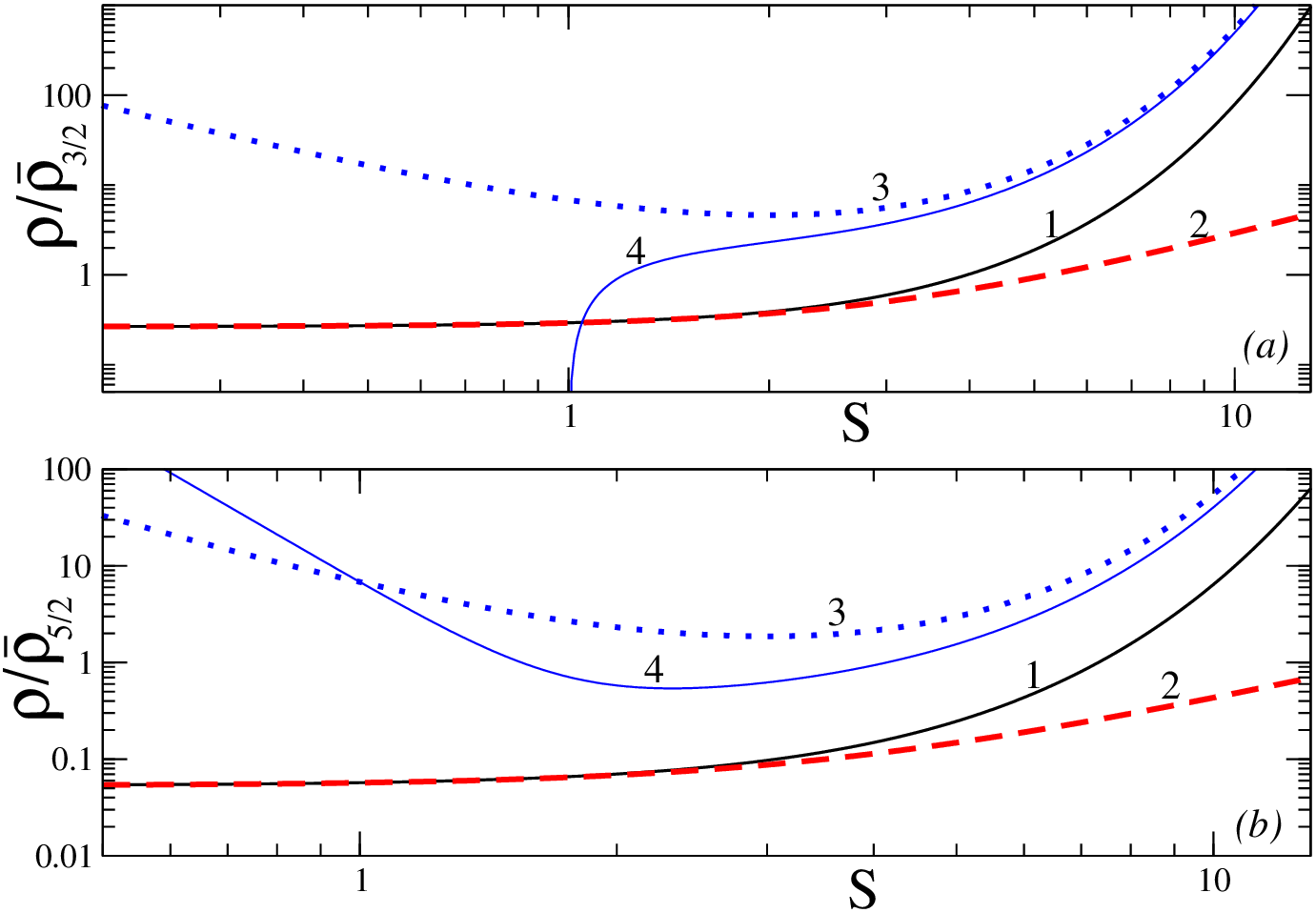}
    \end{center}

\vspace{-0.5cm}
\vskip-3mm\caption{{\small
    Level density $\rho$
    in units of $\overline{\rho}_\nu$, (a) and (b)
    for $\nu=$
    $3/2$ and
    $5/2$ (solid ``1''), respectively,
    are
    shown as
    functions of the entropy $S$ in 
    different approximations:
    1) $S \ll 1$ (dashed ``2'') Eq.~(\ref{den0gen}),
    and 2) $S \gg 1$
    (dots and thin solid),~ 
        Eq.~(\ref{rhoasgen}), 
     ``3'' and ``4'',   are 
    expansions over $1/S$ 
    up to zero and first [in (a)]
   or second [in (b)]
    order terms in 
    Eq.\ (\ref{rhoasgen}), respectively.
}}
\label{fig1}
\end{figure}

In the opposite case (ii) of the dominating 
    shell-corrections contributions,
$\xi \gg 1$, see Eq.~(\ref{ratio}),  assuming  
a relatively
large (heat) excitation
energy with respect to the collective rotational energy,
one obtains Eq.~(\ref{Q}) for the Jacobian.
The  inverse Laplace integrand 
requires  
the additional power of 
$\beta$. Therefore, after transformation
of the variable, $\beta=1/\tau$,
and taking explicitly the inverse Laplace transform
we arrive at the same expressions of Eq.~(\ref{denbes}) but
with $\nu=5/2$ (MMA2 approximation),
    \begin{equation}\label{ld52}
      \rho_{\tt{MMA2}}(E,A)=\overline{\rho}^{}_{5/2} S^{-5/2}I_{5/2}(S)~,\quad
     \overline{\rho}_{5/2}=
     4a^2\left(\pi/6\overline{\xi}\right)^{1/2}~,\qquad (ii);
    \end{equation}
    where 
\begin{equation}\label{xibdE}
\overline{\xi}=\beta^2 \xi
\approx
\frac{4\pi^6 A^{1/3}\mathcal{E}_{\rm sh}}{3\lambda^2}~, 
  \end{equation}
see Eq.~(\ref{ratio}) for $\mathcal{E}_{\rm sh}$.

We will specify MMA2 approach (ii), Eq.~(\ref{ld52}),
as MMA2a one for  
$\mathcal{E}_{\rm sh}$ taken in 
  Eq.~(\ref{xibdE}) from
Ref.~\cite{MSIS12}.
For the TF approximation to the coefficient $\overline{\rho}_{5/2}$ 
within the
 same case (ii) 
one finds (MMA2b approximation) \cite{PRC}
\begin{equation}\label{ld52-2b}
  \rho^{}_{\tt{MMA2b}}(S)=\overline{\rho}^{(2b)}_{5/2}~S^{-5/2}I_{5/2}(S), \qquad
  \overline{\rho}^{(2b)}_{5/2}\approx 2\sqrt{2/\pi}~\lambda
 ~ a^2~,~~~~(ii).  
\end{equation}
For derivations of the coefficient, $\overline{\rho}^{(2b)}_{5/2}$, we assumed in
    Eq.~(\ref{ld52}) for $\overline{\rho}_{5/2}$ that the magnitude
of the  relative shell corrections
    $\mathcal{E}_{\rm sh}$, $\overline{\xi}\propto \mathcal{E}_{\rm sh}$;
    see Eq.~(\ref{xibdE}),
    is  extremely 
small but their derivatives yield large
contributions through the 
level density derivatives  $g^{\prime\prime}(\lambda)$ [Eqs.~(\ref{d2Edl2})],
$g \propto A/\lambda$, 
as in the TF approach.
    In this case for the  spin-projection-dependent level density
    $\rho(E,A,M)$, one obtains the same expressions of Eq.~(\ref{denbes})
    but with
   $\nu=3$ and $\overline{\rho}_3 \approx (4\sqrt{2}a^3 \hbar/\pi^2)
    \left(\lambda^3/A \Theta\right)^{1/2}$ in Eq.~(\ref{denbes}),
    where  $\Theta$ is the moment of inertia, 
   which  is approximated by the TF
   nuclear moment of inertia $\Theta^{}_{\rm TF}$,
   $\Theta \approx \Theta^{}_{\rm TF}$,  see Ref.~\cite{PRC}.
    The values of $\nu$ in the case (ii) correspond to $\nu-1$
    of the case (i).  Subscript $\nu $ of the Bessel functions in
    Eq.~(\ref{denbes})
    is increased by 1/2 for isotopically asymmetric neutron-proton systems.

    The 
 asymptote for
    large entropy $S$
    is given by
 \begin{equation} 
 f_\nu(S) =\frac{\exp(S)}{S^{\nu}\sqrt{2\pi S}}\left[1+\frac{1-4\nu^2}{8S}
    +\mbox{O}\left(\frac{1}{S^2}\right)\right].
 \label{rhoasgen}
\end{equation}
 This approximation at zero order in
     expansion over $1/S$ is identical to that
     obtained directly from Eq.~(\ref{rhoE1}) by the SPM over all
     variables. At small entropy, $S \ll 1$, one obtains
     also from Eq.~(\ref{denbes})
     the finite combinatoric power
expansion \cite{St58,Er60,Ig72}:
\begin{equation}
    f_\nu(S)=
  \frac{2^{-\nu}}{\Gamma(\nu+1)}\left[1+\frac{S^2}{4(\nu+1)}+
    \mbox{O}\left(S^4\right)\right],
\label{den0gen}
\end{equation}
where $\Gamma(x)$ is the gamma function.
    This expansion over powers of 
    $S^2 \propto U$ 
    is similar to 
    that of  the 
    ``constant temperature model'' 
    \cite{GC65,ZS16,ZK18,ZH19},
used often for the level density calculations  at
small excitation
energies $U$, but here we have it without
free  fitting parameters.   Moreover, we do not need to match 
    asymptotes
for large, Eq.~(\ref{rhoasgen}), and small, Eq.~(\ref{den0gen}), excitation energies $U$
because they are connected analytically by the Bessel function, Eq.~(\ref{denbes}),
and therefore, have no extra free fitting parameters for their matching
in the MMA. 
Using the asymptotes of Eq.\ (\ref{rhoasgen})  in the case (i),
    Eq.\ (\ref{ld32}),
    for large $S$, one arrives at the Bethe expression for the level density
    \cite{Be36,Er60,BM67},
  \begin{equation}
  \rho^{}_{\rm FG}(E,A)= \frac{\exp\left(S\right)}{\sqrt{48}U}~.
\label{FG}
  \end{equation}
This asymptotic expression is obviously divergent at $U\rightarrow 0$, in
contrast to the finite MMA limit (\ref{den0gen}) for the level density,
Eq.~(\ref{denbes}).

\begin{figure}
\begin{center}
\includegraphics[width=0.7\columnwidth]{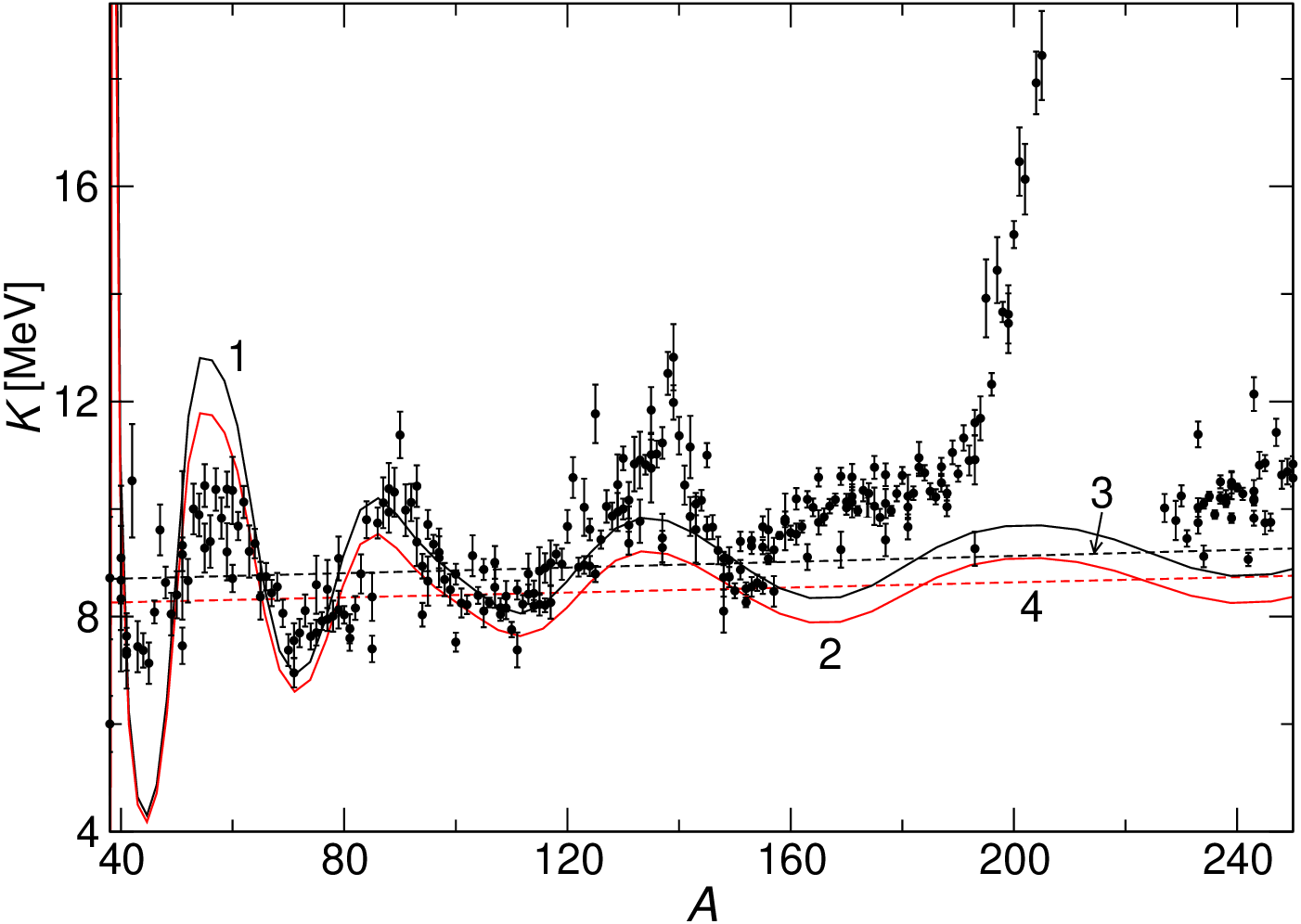}
\end{center}

\vskip-5mm\caption{{\small 
    The inverse level-density parameter $K=A/a$ (solids ``1'' for SKM$^\ast$
   and ``2'' for KDE0v1 forces) is shown as function
    of the particle number $A$. 
    The smooth
    part in  the ETF approach is taken from Ref.~\cite{KS18}
    for these two versions of the Skyrme
    forces SKM$^\ast$ (``3'' dashed) 
    and KDE0v1 (``4'' dashed). The solid
       oscillating curves are
    obtained
    by using the semiclassical POT approximation, Eq.~(\ref{dgsph}),
    for the level density shell corrections, Eq.~(\ref{avdennp}), at Gauss width averaging
    parameter $\gamma=0.3$,
    and dashed curves ``3'' and
    ``4'' for a smooth part, both
    including the effective mass contribution.
    Experimental values,  
    shown by solid points, are taken from Ref.~\cite{EB09}.}}
\label{fig2}
\end{figure}

 Notice that the MMA1 approximation for the level density,
  $\rho(E,A)$,
  Eq.~(\ref{ld32}),
  can be applied also for
  large excitation energies, $U$, with respect to the
  collective rotational excitations,
  as the 
  FG approximation 
  if one can neglect shell effects, $\xi\ll 1$.
  Thus, the level density $\rho(E,A)$ in the case (i),
      Eq.~(\ref{ld32}),
      has a wider range of the
      applicability over the excitation energy variable
  $U $ than the MMA2 case (ii).
  The MMA2 approach has, however, another advantage 
  of describing the important shell structure effects. 
     We emphasize also that the main effects of
  the interparticle
  interaction, statistically averaged over particle numbers,
  beyond the shell correction of the mean field within the Strutinsky's
  shell correction 
      method, was taken into account by the extended Thomas-Fermi
  components of MMA expression
  (\ref{denbes}) for the level density, $\rho(E,A)$. These components
  are given by 
  the extended Thomas-Fermi potential,
  $\Omega^{}_{\rm \tt{ETF}}$, Eq.~(\ref{TFpotF}),
  and the level-density parameter, $a^{}_{\rm \tt{ETF}}$, counterparts of
  the corresponding
  total quantities,
  Eqs.~(\ref{OmFnp}) and (\ref{aSCM}).

\section{Results and discussions}
\label{resdisc}

In Fig.~\ref{fig1}
we show
the level density dependence
$\rho(S)$ [Eq.~(\ref{denbes}), 
  for $\nu=3/2$ in Fig.~\ref{fig1} (a) and $\nu=5/2$ in
  Fig.~\ref{fig1} (b) panels]  on the
entropy variable $S$
and its different asymptotes. In this figure, the results
  of
a small [$S\ll 1$, Eq.~(\ref{den0gen})] and
large [$S\gg 1$, Eq.~(\ref{rhoasgen})] entropy $S$ behavior 
are presented. For large $S\gg 1 $
we neglected the corrections of the inverse power expansion of the
pre-exponent factor
in square brackets of Eq.\  (\ref{rhoasgen}), lines ``3''.
We also
took into account
the 
corrections of the
first [$\nu=3/2$, (a)] and up to second [$\nu=5/2$, (b)]
    order in $1/S$ 
to show  
their slow convergence
to the exact MMA result ``1'' [Eq.~(\ref{denbes})].
It is interesting to find almost a parallel
constant shift of the simplest, $\rho \propto \exp(S)/S^{\nu}$,
SPM 
asymptotic approximation at
large $S$ (dots ``3'')
with respect to the solid curve of 
the exact 
MMA result (\ref{denbes}).
This 
may clarify one of the phenomenological
models, e.g., the 
back-shifted
Fermi-gas 
(BSFG) model for the level density \cite{DS73,So90,EB09}.

Fig.~\ref{fig2} shows 
the inverse level-density parameter, $K=A/a$, with $a$
    of Eq.~(\ref{a}) 
as function of the particle number $A$ in  
    the semiclassical POT
approximation (see \ref{appA}). 
The result of these
calculations are largely in a qualitative agreement with 
\noindent
the 
 recent experimental data \cite{EB09}, which, as compared
to Ref.~\cite{St72},   
included in the analysis of many other excited
  nuclei and different
    reactions
    with nuclear excitation energies being significantly smaller
    than the neutron
    separation energy.
The sets with reliable
  completeness of levels in the limited energy range below
the neutron binding energy   
were selected for each nucleus, and neutron
resonance level densities were included in the analysis.
We added the smooth self-consistent ETF values $\tilde{a}$ of $a$
[Eq.~(\ref{aSCM})]
for the KDE0v1 \cite{ss1} and
the SkM$^\ast$ \cite{BG85} Skyrme forces
 from
 Refs.~\cite{KS18,KS20} to their 
   shell corrections, 
 $\delta g(\lambda)$,
   through the total 
   level density
$g(\lambda)$ of Eq.~(\ref{a}).
Its oscillating component $\delta g(\lambda)$ was
approximated by the analytical POT trace formula \cite{BB72,SM76,BB03} for the
infinitely deep spherical square-well potential; see
Eqs.~(\ref{avdennp}) and
(\ref{dgsph}).
   This formula  reproduces, almost
   identically,  the
   quantum results of the shell correction method for the 
   level density $g$ 
  in the same potential \cite{BB03}.  
The major closed shell nuclei are clearly recognized 
in Fig.\ \ref{fig2} by the maxima of $~K(A)~$, which correspond
to minima of the level density parameter $a$, or
the oscillating level density
component $\delta g(\lambda)$,
see 
Eqs.~(\ref{a}) and (\ref{dgsph}).
The relationship between the chemical potential $\lambda$, through $k^{}_FR$,
where $\hbar k^{}_F=(2m\lambda)^{1/2}\approx (2m\varepsilon^{}_F)^{1/2}$ is
the Fermi momentum, $\varepsilon^{}_F$ is the Fermi energy, and $m$
the nucleon   

\vspace{-0.4cm}
\begin{figure}
  \begin{center}
   \includegraphics[width=10.0cm]{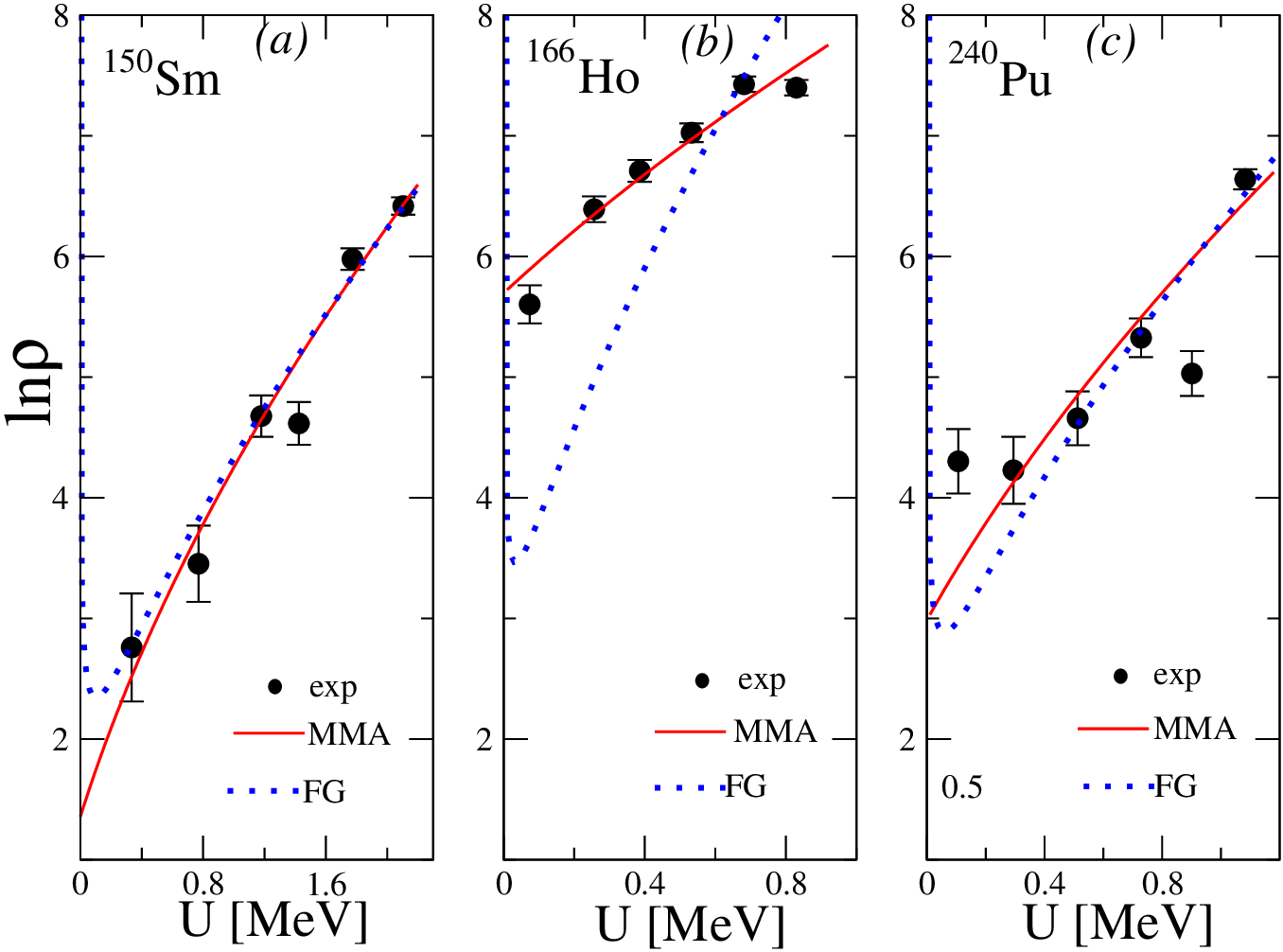} 
    \end{center}
  
  \vskip-5mm\caption{{\small
  Level density, $\mbox{ln}\rho(E,A)$, obtained for
    low energy states in nuclei $^{150}$Sm (a), $^{166}$Ho (b), and $^{240}$Pu (c)
    in the MMA for the smallest $\sigma$, Eq.~(\ref{chi}), in
    each nucleus among
    all MMA approximations,
    MMA1 [Eq.~(\ref{ld32})], MMA2b [Eq.~(\ref{ld52-2b})], and MMA2a
    [Eq.~(\ref{ld52})] with the
    realistic relative shell-energy 
    corrections, $\mathcal{E}_{\rm sh}$
    from Ref.~\cite{MSIS12}),
    respectively;
    the 
    chemical potential is constant, 
$\lambda=40$ MeV, independent of particle numbers. 
    The MMA is  compared with the standard Fermi gas 
    formula,
    Eq.~(\ref{FG}).
    The relative shell corrections, $\mathcal{E}_{\rm sh}$, are taken
    from Ref.~\cite{MSIS12},
    $\mathcal{E}_{\rm sh}=0.2, 0.5$ and $0.7$ for the
    $^{150}$Sm (a), $^{166}$Ho (b) and $^{240}$Pu (c)
    nuclei,
    respectively.
    Experimental dots 
    are obtained 
    from the 
    excited 
    states spectrum of these nuclei
    (see 
    Ref.~\cite{ENSDFdatabase})
    by using the sample method, see text.
}}
\label{fig3}
\end{figure}

\vspace{0.4cm}
\noindent
mass. The particle numbers $A$ 
for this potential \cite{SM76,BB03} is given by the second equation of
Eq.~(\ref{conseq}). In 
the transformation of
$k^{}_FR$ (or $\lambda$) to $A$, Eq.~(\ref{conseq}),
one can conveniently use the quantum shell-correction level density \cite{BD72}.
The Gaussian averaging width of the oscillating level density $\delta g$
in
Fig.~\ref{fig2} [see Eq.~(\ref{avdennp})] is the same,
$\gamma=0.3$, as that adopted in previous POT calculations \cite{MY11}.
It
corresponds to the dimensional Gaussian width 
$\Gamma\approx  2\gamma \lambda/k^{}_FR\approx 3$ MeV
($\lambda =40$ MeV, $R=r_0 A^{1/3}$, $r^{}_0=1.14$ fm, and
$A=100-200$).
In calculations results presented in 
 Fig.~\ref{fig2}, 
only short planar POs yield the
main major-shell contributions into the PO sum,
Eqs.~(\ref{avdennp}) and (\ref{dgsph}),
for the oscillating level density 
in Eq.~(\ref{a}).
Mean value of
oscillating $K(A)$ 
in Fig.~\ref{fig2}  is about
 8 MeV, 
as predicted 
in Ref.~\cite{Sh92}. This is in accordance with  
the ETF (SkM$^\ast$ or KDE0v1) value, associated with the
effective mass $m^\ast$.
As shown in Ref.~\cite{KS18}, the effect of the effective mass
$m^\ast$ on the inverse
level density parameter $K$
is
strong, decreasing of $K$
by a factor of about 2, 
that leads approximately to mean values of the experimental data \cite{EB09}.
However, we should not expect that for the infinitely deep spherical square-well
potential. The positions of the maxima [minima of
$a$, i.e., of the 
  single-particle level density $\delta g(\varepsilon)$
  at $\varepsilon=\lambda$, related
to magic nuclei in that potential] can not be correctly reproduced  in
such shell-correction
calculations,
because
of
neglecting the
spin-orbit interaction. As shown in Refs.~\cite{SM77,MY11},
in order to reproduce the experimental value of $K$
for the magic nucleus $^{240}$Pu in the semiclassical calculations, we should
shift the curves $K(A)$ along the $A$ axis [through $k^{}_FR$,
Eq.~(\ref{conseq})].  Therefore, we shifted the semiclassical curves in
Fig.~\ref{fig2} with
about $\Delta A=20$
along the particle number $A$ axis.
This shift is of the order 
of the period
$A_{\rm sh}$, Eq.~(\ref{dA}), 
related to the distance $D_{\rm sh}$, Eq.~(\ref{periodenp}),
between major shells near the Fermi surface in the particle number 
  variable
$A$.
  According to the POT estimations  (see \ref{appA}
  and Ref.~\cite{SM76}) for
the period $D_{\rm sh}$ and TF level density
    $g^{}_{\rm TF}\sim A/\lambda$,
one finds  Eq.~(\ref{dA}) for the particle number period of the
major shell structure, $A_{\rm sh}$.
 For $A=100-200$, one
 obtains $A_{\rm sh} \approx 
 20-30$, which
 is of the order of the realistic period
 of the nuclear major shell structure; see also
   Ref.~\cite{MSIS12}.
The position of a maximum
over the particle number variable, as related to the value of $k^{}_FR$,
is the only one parameter needed to adjust the semiclassical
solid curves in Fig.~\ref{fig2} and compare with the experimental
data.
This  is similar to the discussions in Ref.~\cite{MY11} where
the magic number for $^{240}$Pu was obtained
semiclassically by using a similar shift.
Therefore, three minima of the 
level density shell corrections
$\delta g$ for the major shell closures 
in the semiclassical calculations
at $A\approx 45-150$ shown in Fig.~\ref{fig2}
correspond to  the 
maxima of the inverse level density parameter, $K\propto 1/a$,
experimentally obtained in Ref.~\cite{EB09}.
In spite of 
very simple explicitly given analytical formulas
(\ref{avdennp}) and (\ref{dgsph})
(Refs.~\cite{BB72,SM76,BB03}) for the 
level-density shell corrections in the spherical cavity, one obtains largely
good agreement of the semiclassical approximation, which is almost 
identical to the quantum 
result  of the shell-correction method for the same cavity,
with the  experimental data in this range of nuclear
particle numbers $A$.
The magnitudes   of
periods
and amplitudes 
for the oscillations of $K(A)$ are basically in good agreement
with experimental data.
However, there is a discrepancy between experimental and theoretical
results for $K(A)$ in the range of particle numbers $A\approx 150-240$.
One of possible reasons is that 
different
approximations for the statistical level density $\rho$ are used
in fitting
    procedure to obtain
    the experimental prediction of $K$ with respect to those of
    the
    MMA approach (\ref{denbes}).
    Experimental data for $K$ are in good agreement with those  of neutron
    resonances, see Refs.~\cite{St72,KS18}, which are dominating
    in the results of the 
    calculations of Ref.~\cite{EB09}.   
The specific reason for the discrepancy might be that the
level density parameter $a$ (or $K$)
was obtained by
 three free parameter 
fit  of these
    experimental data
    to the
    level density 
    within the  BSFG and constant
    temperature models.
     Another reason is that the pairing effects which, as well known,
      are strong
    in $^{208}$Pb should be accounted for such nuclei with
    a large pairing gap in the excited spectra, see Ref.~\cite{PRC}.

In Fig.\ \ref{fig3} 
we present results of 
the MMA and standard FG 
approximations
to the statistical level density $\rho(E,A)$ (in logarithms) as functions of the
excitation energy $U$ versus the experimental
data. They are calculated  
 employing the inverse level density parameter $K$
deduced from 
fits to experimental data 
  on the excited energies in a low energy-states (LES) range
for several nuclei.
The experimental data used for the statistical level density
    $\rho(E,A)$
are obtained 
from Ref.~\cite{ENSDFdatabase}
for the excitation energies $U$ and spins 
with accounting for spin degeneracies by using the sample method
 \cite{So90,PRC,IJMPE}. In Fig.~\ref{fig3} we present the two opposite situations
concerning the states distributions as functions of the excitation energy $U$.
We 
show results for $^{150}$Sm in Fig.~\ref{fig3}(a) 
for the case of almost no states with extremely low
excitation energies, 
similarly as for $^{208}$Pb where there is no such levels at all. Only a few
levels in $^{150}$Sm can be found
at $U \siml 1$ MeV, which yield
entropies $S \siml 1$.
For $^{166}$Ho 
in Fig.~\ref{fig3}(b), one finds the opposite situation when
there is a lot of such LESs. An intermediate number of LESs is observed, e.g.,
in  heavy 
  $^{240}$Pu
(Fig.~\ref{fig3}(c)). Thus, we present the results for nuclei,
$^{150}$Sm ($^{166}$Ho), and $^{240}$Pu,
from both sides of the desired particle number interval $A\approx 150-240$.
In Fig.~\ref{fig3}, we present  
the results of the MMA,  Eqs.~(\ref{ld32}), (\ref{ld52-2b}) and
(\ref{ld52}), at the minimal 
control relative error-dispersion parameters of the least mean square
fit:
\begin{equation}\label{chi}
\sigma^2=\frac{\chi^2}{\aleph-1}, \qquad \chi^2 =\sum_i
\frac{(y(U_i)-y^{\rm exp}_i)^2}{(\Delta y_i)^2},\qquad y=\ln\rho~,
  \end{equation}
where $\Delta y_i 
\sim 1/\sqrt{N_i}$, 
  $N_i$  is the number of the excited states in the i$th$ sample, and $\aleph$
is the sample number, $i=1, 2, ..., \aleph$.
The number $N_i$ is large, according to the statistical condition of the
sample method.
 The errors $\Delta y_i$
in 
statistical distributions of
the 
excited states of the available experimental data over
samples 
were evaluated as units of the theoretical versus
experimental differences
$y(U_i)-y^{\rm exp}_i$.
    We  determine $\sigma$, Eq.~(\ref{chi}), at the minimum of
    $\chi^2$ over the unique parameter, $K=K_{\rm min}$,
    having a definite physical meaning as the
    inverse level density parameter $K$  (see Fig.~\ref{fig2}).
      Then, we may compare the $\sigma$ value of
    several different  MMA 
    approaches, which were found independently of 
    the data,
  under certain statistical conditions mentioned above.
    For this aim we are interested in
  relative values of the
  $\sigma(K_{\rm min})$ for different compared approximations
  rather than their absolute values.
  The MMA  results for the minimal values of $\sigma$,
  Eq.~(\ref{chi}), are shown
  in plots of Fig.~\ref{fig3} by red solid lines as best, among the
  MMA, approaches. As seen from these plots, they 
  agree well with the experimental data.  The results of our calculations
    are almost independent of
    the sample number, $\aleph=5-7$, which plays the same role as an
    averaging parameter on
    the plateau condition, as in the Strutinsky averaging procedure
    \cite{BD72}.

In 
all panels
of Fig.~\ref{fig3}, one can see the divergence
of the FG
level density,  see Eq.~(\ref{FG}), 
near the zero excitation energy
$U\rightarrow 0$, in contrast to the finite combinatoric MMA expressions
 (Refs.~\cite{St58,Er60,Ig72}) of
Eq.~(\ref{den0gen})
in this limit,
see 
Eq.~(\ref{denbes}).
We do not use 
empiric  fitting parameters
of the BSFG, in particular, spin cut-off parameter 
and those 
of the empiric constant temperature model
\cite{EB09}. 
As an advantage,  as mentioned above, one has
only one  fitting
parameter $K$
which has a certain physical meaning as the inverse level density parameter.
The variations of $K$ are related, e.g., to those of the mean field parameters
through Eq.~(\ref{a}), see \ref{appA}.
All the MMA densities $\rho(E,A)$ 
do not depend 
on the cut-off spin factor
and moment of inertia 
because of summation (integrations) over all spins 
(with accounting for the spin degeneracy factor) which could
appear in the spectrum.
The MMA results for $^{150}$Sm (MMA1, case (i)),  $^{166}$Ho (MMA2b, case (ii)),
and $^{240}$Pu
(MMA2a, case (ii))
are compared with the results of the FG approximation
and the experimental data in Fig.~\ref{fig3}. Obviously,
  much
  better agreement with data is obtained for
  $^{166}$Ho (Fig.~\ref{fig3}(b)) with a lot of states in the very LESs range.
The MMA1  results are 
mainly agreed with experimental data 
for $^{150}$Sm (Figure \ref{fig3}(a)),  as well the FG approach,
for which one has the opposite
situation. For  $^{240}$Pu (Figure \ref{fig3}(c)), one obtains within the
MMA  (MMA2a) a better agreement with the experimental data at LESs 
than with the results of the FG formula (\ref{FG}).
 In any case, in the MMA, we removed the
 divergence of the statistical level density, $\rho(E,A)$,
 at small excitation energies $U$.
 Notice that the shell corrections effects measured by $\mathcal{E}_{\rm sh}$,
  Eq.~(\ref{ratio}), increase going
  from (a) to (c)
   panels of Fig.~\ref{fig3}, 
    largely
  in 
  agreement with the derivations of the MMA approaches.

In line with the results of Refs.~\cite{ZS16,ZH19},  the MMA 
results for $K$ can be significantly
larger than the FG ones, and those 
obtained mainly for the neutron resonances.
For instance, for very low excitation energy states
in  $^{166}$Ho,  for the MMA2b approach
  one obtains
  $K\approx 17.6$ MeV ($\sigma\approx 1.9$) while for the FG approach one finds
  $K\approx 5.5$ MeV ($\sigma\approx 11.1$), cf.\ with the experimental data
  for neutron resonances, $K=9.7$ MeV in the same nucleus,
  see Fig.~\ref{fig2} and Ref.~\cite{EB09}. 
For MMA1, one finds $K$ of the same order of magnitude as
that of the FG approach, in contrast to the  MMA2b.
The MMA1 ($K\approx 5.4$ MeV, $\sigma\approx 12.1$ for $^{166}$Ho)
and FG 
values of $K$ are mostly 
close to  those of 
neutron resonances 
in order of
magnitude (Fig.~\ref{fig2}).
For the FG case it is obviously because
 they occur at large
excitation energies $U$.
 Notice that in all our calculations of the statistical level density,
$\rho(E,A)$, we did not use a popular assumption of small spins at large
excitation energies which is valid for the neutron resonances.
Large deformations, neutron-proton asymmetry and pairing correlations
  \cite{Er60,Ig83,So90,AB00,AB03,ZK18,ZH19} of
  the rare earth and
      actinide nuclei should be also
      taken into account \cite{PRC,IJMPE}
      to improve the comparison with experimental data.

      \section{Conclusions}
      \label{concl}
    
We derived the statistical level density $\rho(S)$ as function of the
entropy $S$
within the micro-macroscopic 
approximation using the mixed micro- and grand-canonical ensembles beyond
the standard saddle point method of the Fermi gas model.
This function can be applied for small and, relatively, large entropies
$S$ or excitation energies
$U$ of a nucleus. For a large entropy (excitation energy), one obtains
the 
exponential asymptotes of the  standard 
Fermi-gas model, however, with 
the significant inverse $1/S$ power corrections.
For small $S$ one finds the usual finite combinatoric
expansion in powers of $S^2 \propto U$. Functionally, the
results of the MMA
  approach at
linear approximation in $S^2$
    expansion, and at small excitation energies,
     coincide with those of the empiric
    constant ``temperature'' model, but without using 
    free fitting parameters. Thus,
   the MMA unifies the
    commonly accepted  
    Fermi-gas approximation, for large entropies $S$,
    with the empiric constant temperature model 
    for 
    small entropies $S$, respectively, in line 
  with  the suggestions in Refs.~\cite{GC65,ZK18,ZH19}.
The MMA results
    clearly manifest an 
advantage
over the standard saddle point 
  approaches, for the case of low excitation energies,
because of 
no divergences of the MMA in the limit of
small
excitation energies, in contrast to the  
 FG asymptote.
Another advantage takes place 
for nuclei which have much more states in the very LES range.
The values of the inverse
level density
parameter K 
are compared with experimental data for LESs which are significantly below
 neutron resonances 
in 
nuclear spectra of 
several nuclei. 
The MMA 
results with only one physical parameter in the least mean-square
fits - inverse level density parameter $K$ -
 are usually the better the larger number of the extremely low energy states,
certainly much better
than for the FG model in this case.  The 
MMA values
of the inverse level density parameter $K$
for LESs 
can be significantly
larger than those of the neutron resonances within the FG model.
 Major shell oscillations
of the inverse level density
parameter $K$
are compared with modern 
experimental data, basically 
for neutron resonances.
We found qualitatively good
agreement between semiclassical POT and quantum-mechanical results with
experimental data on the inverse level
density parameter $K$  for neutron resonances,
after overall shift of all
$K(A)$ curves between particle numbers $A\approx 45-150$ by only one
parameter because of the spin-orbit interaction.

We have found 
 significant shell effects in the MMA level density for the
nuclear low excitation states (LES) range
within the semiclassical periodic-orbit theory (POT).
Therefore, a reasonable description of the LES experimental data for the
statistical averaged level density
obtained by the sample method
within the MMA  approach with the help of the semiclassical POT was achieved
 for several typical nuclei in the $A\approx 150-240$ range.
We emphasize the
importance of the shell 
effects in these
calculations. We obtained values of $K$
that are significantly 
larger 
than those obtained
for 
neutron resonances,  due mainly to 
accounting for  
the shell effects. We show that the semiclassical periodic orbit theory is
helpful in the LES range
  for analytical description of the level density
   and energy shell corrections.
  They are taken into account
   in the linear approximation up to small corrections due to the
   {\it residual} interaction beyond
   the mean field and extended Thomas-Fermi approximation
   within the shell-correction method,
   see Refs.~\cite{BD72,BK72}. The main part of the
    interparticle interaction is described in terms of 
   the extended Thomas-Fermi counterparts of the
   statistically averaged nuclear potential, and in particular,
   of the
   level density parameter.
We obtained 
values of the inverse level
density parameter $K$ for the
LES range which are essentially different
from those 
of  neutron resonances. 
We found 
significantly 
larger $K$ values,
than those
for neutron resonances 
because of accounting for  
the shell effects.

As perspectives,
the neutron-proton asymmetries, 
large nuclear angular momenta and
deformations, 
as well as pairing correlations,  will be 
taken into account in future work to significantly improve the comparison of the
theoretical evaluations of the level density parameter 
with
experimental data  
below the neutron resonances.  Following the ideas of Refs.~\cite{Bj74,BM75,Ju98,Gr13,Gr19}
and using the MMA results of Ref.~\cite{PRC},
we are going to derive the spin-dependent level density for collective rotations of
the axially symmetric deformed nuclei.
We will study the enhancement factors of the level density
due to
their collective rotations and vibrations within the analytical MMA in
the adiabatic approximation.
Our approach can be applied also to the statistical analysis of
the experimental
data on collective nuclear states obtained in several nuclear reactions.
The MMA  approach with essential shell effects can be applied also
for metallic
  clusters and quantum dots, and therefore,
  in solving several problems 
 in nuclear astrophysics.

\vspace{1.0cm}
\bigskip
\centerline{{\bf Acknowledgement}}
\bigskip

The authors gratefully acknowledge 
D.\ Bucurescu,
  R.K.\ Bhaduri, M.\ Brack,
 A.N.\ Gorbachenko, and V.A.\ Plujko for creative discussions. 
This work was supported in part by the  budget program
"Support for the development
of priority areas of scientific researches",  the project of the
Academy of Sciences of Ukraine,
Code 6541230, No 0122U000848.
S.\ Shlomo is supported in part
by the US
Department of Energy under Grant no. DE-FG03-93ER-40773.

\appendix

\section{
The semiclassical POT}
\label{appA}

The level density shell corrections for 
 a finite nucleon one-component system describing in the
  mean field approximation
can be presented analytically
within the periodic orbit theory (POT)
in terms of  the sum over classical periodic orbits (PO) \cite{SM76,BB03,MY11},
\begin{equation}
  \label{goscsemnp}
\delta g^{}_{\rm scl}(\varepsilon)=
\sum^{}_{\rm PO}g^{}_{\rm PO}(\varepsilon),\quad
g^{}_{\rm PO}(\varepsilon)=\mathcal{A}_{\rm PO}(\varepsilon)
~\cos\left[\frac{1}{\hbar}\mathcal{S}_{\rm PO}(\varepsilon)-
\frac{\pi}{2} \mu_{\rm PO}
-\phi_0\right].
\end{equation}
Here $\mathcal{S}_{\rm PO}(\varepsilon)$ is the classical action along the
PO in the 
potential well of 
radius,
$R=r_0A^{1/3}$ ($r^{}_0\approx 1.1$ fm),
$\mu_{\rm PO}$ is the so called Maslov index determined by
the catastrophe points (turning and caustic points) along the PO, and
$\phi_0$ is an additional shift of the phase coming from the dimension
of the problem and degeneracy of the POs. The amplitude
$\mathcal{A}_{\rm PO}(\varepsilon)$, as the action
$\mathcal{S}_{\rm PO}(\varepsilon)$,
are smooth functions of
the  
 single-particle energy
$\varepsilon$, but, in addition, depend on the PO stability factors.
 The oscillating  component,
    $\delta g(\varepsilon)$, Eq.~(\ref{goscsemnp}), was
approximated, with good accuracy, by the analytical POT trace
formula \cite{BB72,SM76,BB03} for the
infinitely deep spherical square-well potential, 
\begin{equation}\label{dgsph}
\delta g^{}_{\rm scl}(kR)=\frac{2mR^2 d_s}{\hbar^2}\left[\sqrt{\frac{kR}{\pi}}
  \sum\limits_{p,t}^{}(-1)^t
  \sin\left(2 \varphi_{pt}\right) \sqrt{\frac{\sin(\varphi_{pt})}{p}}
  \sin\left(kL_{pt} -3p\frac{\pi}{2}\ +\frac{3\pi}{4}\right)
- \sum\limits^{}_t\frac{\sin(4tkR)}{2\pi t}\right],
  \end{equation}
where  $d_s$ is the spin (spin-isospin)
degeneracy, $k=\sqrt{2m\varepsilon/\hbar^2}$ is the wave number,
 $L_{pt}=2pR~\sin(\varphi^{}_{pt})$ is the PO length, $R$ is
 the radius
   of the spherical cavity, $\varphi^{}_{pt}=\pi t/p$, and
  $p$ and $t$ are the numbers of turning points and rotations around the center
   of the PO (winding number) in the spherical 
       cavity, respectively. 
   The first and second terms in square brackets of
   Eq.~(\ref{dgsph}) are
   the contributions  of families of the planar orbits and
   diameters.
      The Gaussian local averaging of the level density shell
       correction, 
          $\delta g^{}_{\rm scl}(\varepsilon)$, 
      with a width
      $\Gamma$, over the
    single-particle energy spectrum
  $\varepsilon^{}_i$ near the Fermi surface $\varepsilon^{}_F$,
   for $\Gamma $ smaller than a distance between major shells,
  $\mathcal{D}_{\rm sh}$, can be done
  analytically \cite{SM76,BB03,MY11}. Similarly, one can average
  $\delta g^{}_{\rm scl}(kR)$ for cavity potentials over $kR$ variable with
  dimensionless $\gamma \approx  \Gamma k^{}_FR /2\lambda $,
  having
  \begin{equation}
    \label{avdennp}
  \delta g^{(\gamma)}_{\rm scl}(kR)\cong \sum^{}_{\rm PO}g^{}_{\rm PO}(kR)~
\exp\left[-\left(\frac{\gamma L^{}_{\rm PO}}{2~R}\right)^2\right]~,  
  \end{equation}
  where $L_{\rm PO}$ is the length of the PO in a given  
cavity potential well, e.g.\ $L_{pt}$ for the PO$(p,t)$ in the infinite
  square well potential.

  Using the 
 MMA approximation for calculations of the potential $\Omega$,  one can express
it in terms of 
 the 
level density $g(\varepsilon)$, as
\begin{equation}\label{OmFnp}
\Omega\approx -\beta^{-1} 
\int\limits_0^\infty\d\varepsilon~
g(\varepsilon)~
\ln\left\{1+\exp\left[\beta\left(\lambda-
  \varepsilon\right)\right]\right\}~.
\end{equation}
 The  
level density  $g(\varepsilon)$, within the Strutinsky
 shell-correction method \cite{BD72} 
is a sum of
the smooth, $\tilde{g}(\varepsilon)$, 
and oscillating shell,
$\delta g(\varepsilon)$, 
 components as
\begin{equation}\label{gdecomp}
g(\varepsilon)\cong \tilde{g}(\varepsilon)+
\delta g(\varepsilon)~.
\end{equation}
     The oscillating part $\delta g(\varepsilon)$ is
     averaged over the 
 single-particle energies near the Fermi energy
 using a small Gauss width, $\Gamma\propto \gamma$;
see Eq.~(\ref{avdennp})  for cavities in the POT and
  Ref.~\cite{BD72} for quantum-mechanical case 
 of 
any potential well. 
Within the semiclassical POT \cite{SM76,SM77,BB03,PRC}, 
the smooth and
oscillating parts
of the 
level density $g$ can be approximated, with good accuracy, by
the ETF level density,
$\tilde{g} \approx g_{\rm \tt{ETF}}$,
and the PO  contribution, $\delta g\approx
\delta g_{\rm scl} $ [see 
Eq.~(\ref{goscsemnp})], respectively.
Using 
  the  POT 
decomposition, Eq.~(\ref{gdecomp}),
one finds from Eq.~(\ref{OmFnp})
that
$\Omega=\tilde{\Omega}+\delta \Omega$,
where
$\tilde{\Omega}\approx \Omega_{\rm\tt{ETF}}$ is the smooth
ETF 
component \cite{KM79,KS20}, given by
\begin{equation}\label{TFpotF}
\tilde{\Omega}
=\tilde{E}
-\lambda A
-\frac{\pi^2}{6\beta^2}\tilde{g}(\lambda)~. 
\end{equation}
 The nuclear ETF energy 
 component,
 $\tilde{E}\approx E_{\rm\tt{ETF}}=\int_0^{\lambda}
  \d \varepsilon~\varepsilon~ g^{}_{\rm \tt{ETF}}(\varepsilon)$,
 or the corresponding liquid-drop
energy, is determined by a smooth chemical potential %, and
$\lambda\approx \tilde{\lambda}$ 
 in the
shell correction method.
With the help of the POT \cite{SM76,SM77,BB03,PRC}, one
obtains \cite{KM79,PRC} for the oscillating (shell)
component, $\delta \Omega$,  see Eq.~(\ref{OmFnp}),
\begin{equation}\label{potoscparFnp}
\delta \Omega= -\beta^{-1} 
\int\limits_0^\infty\d\varepsilon~
\delta g(\varepsilon)~
\ln\left\{1+\exp\left[\beta\left(\lambda-
  \varepsilon\right)\right]\right\}
\cong
\delta \Omega_{\rm scl}
=\delta F_{\rm scl}.
\end{equation}
For the
semiclassical free-energy shell correction, $\delta F_{\rm scl}$,
or $\delta \Omega_{\rm scl}$,
we incorporate 
the POT expression \cite{KM79,BB03}: 
\begin{equation}\label{FESCFnp}
\delta F_{\rm scl} \cong \sum^{}_{\rm PO} F_{\rm PO}~,\quad\mbox{with}
\quad
F_{\rm PO}= E_{\rm PO}~
\frac{x_{\rm PO}}{
  \sinh\left(x_{\rm PO}\right)}~,\quad x_{\rm PO}=
\frac{\pi t_{\rm PO}}{\hbar \beta}~,
\end{equation}
where $E_{\rm PO}$ is a PO component of the semiclassical 
shell correction energy,
$\delta E_{\rm scl}$, given by
\begin{equation}\label{dEPO0Fnp}
E_{\rm PO}=\frac{\hbar^2}{(t^{}_{\rm PO})^2}\,
g^{}_{\rm PO}(\lambda)~, \quad \delta E_{\rm scl}=
\sum^{}_{\rm PO}E_{\rm PO}~.
\end{equation}
Here $t^{}_{\rm PO} = {\tt M}~t^{{\tt M}=1}_{\rm PO}(\lambda)$
is the period 
of particle motion
along 
a PO (taking into
account its repetition, or period number ${\tt M}$), 
and
$t^{{\tt M}=1}_{\rm PO}(\lambda)$ is the
period of the particle 
motion along the
primitive 
(${\tt M}=1$) PO in the 
potential well 
    with the 
radius $R=r_0A^{1/3}$.
The period $t^{}_{\rm PO}$ (and $t^{{\tt M}=1}_{\rm PO}$), and
the partial oscillating level density component, $g^{}_{\rm PO}$,
are taken at the chemical potential, $\varepsilon=\lambda$;
see also Eq.~(\ref{goscsemnp}) 
for the semiclassical
level-density shell correction
(Refs.~\cite{SM76,BB03}). The semiclassical expressions,
Eqs.~(\ref{TFpotF}) and (\ref{potoscparFnp}), are valid for a large
relative action, $\mathcal{S}_{\rm PO}/\hbar \sim A^{1/3} \gg 1$.
Then, expanding 
$x_{\rm PO}/\sinh(x_{\rm PO})$,
Eq.~(\ref{FESCFnp}), in the shell correction $\delta \Omega$
[Eqs.~(\ref{potoscparFnp}) and (\ref{FESCFnp})]
in powers of  $1/\beta^2$
up to the quadratic terms, $\propto 1/\beta^2$,  for
the excitation energies much smaller than the chemical potential,
$U \ll \lambda$, see Ref.~\cite{PRC},
one obtains Eq.~(\ref{Ompot}).
 The chemical potential $\lambda$ (or  $\tilde{\lambda}$) 
  is
 the solution of the 
 conservation of particle numbers 
 equation 
 in Eq.~(\ref{conseq}).
The POT shell component of the free energy,
$\delta F_{\rm scl}$,  Eq.~(\ref{FESCFnp}),
is related in the nonthermal and nonrotational limit to the 
     shell correction energy of a cold nucleus,
    $\delta E_{\rm scl}$; see Eq.~(\ref{dEPO0Fnp}) and
    Refs.~\cite{SM76,BB03,MY11}.
Within the POT, $\delta E_{\rm scl}$
is determined, in turn, through Eq.~(\ref{dEPO0Fnp}) by 
the oscillating level density 
$\delta g_{\rm scl}(\varepsilon)$ at the chemical potential,
$\varepsilon=\lambda$;
see Eq.\ (\ref{goscsemnp}).
The chemical potential $\lambda$ can be approximated by the Fermi energy
$\varepsilon^{}_F$ up to  small excitation-energy
corrections ($T\ll \lambda$
for the  saddle point 
value $T=1/\beta^\ast$ if exists). 
  It is determined by the particle-number conservation condition,
  Eq.~(\ref{conseq}),
  where
  $g(\varepsilon)\cong g^{}_{\rm scl}=
  g^{}_{\rm \tt{ETF}} +\delta g^{}_{\rm scl}$
  is the total 
  POT level
density.
One now needs to solve the second equation of Eq.~(\ref{conseq}) to
determine the 
chemical potential $\lambda$ as
   function
  of 
  the particle numbers, $A $, since  
$\lambda$  is needed in Eq.\  (\ref{dEPO0Fnp}) to 
obtain the semiclassical energy shell correction
$\delta E_{\rm scl}$.

For a major shell structure near the  Fermi surface,
$\varepsilon\approx \lambda$,
the POT shell corrections,  $\delta E_{\rm scl}$ [Eq.~ (\ref{dEPO0Fnp})]
and 
$\delta g^{}_{\rm scl}(\lambda)$ 
[Eq.\ (\ref{goscsemnp})] 
are, in fact, approximately proportional to each other.
Indeed, the rapid convergence of the PO sum in Eq.~(\ref{dEPO0Fnp})
is guaranteed by the 
factor in front of the density component $g^{}_{\rm PO}$,
Eq.\ (\ref{goscsemnp}), a factor 
which is inversely proportional to the period
time $t^{}_{\rm PO}(\lambda)$ squared along 
the PO. Therefore, only POs with 
short periods which occupy a 
significant 
phase-space volume near the Fermi surface will contribute.
These orbits are responsible for the
major shell structure, that is related to a Gaussian averaging 
    width,
$\Gamma\approx \Gamma_{\rm sh}$, which is much larger
    than the distance between neighboring 
 single-particle states but much smaller
than the distance
$D_{\rm sh} $ between major shells near the Fermi surface.
Eq.~(\ref{avdennp}) for the averaged 
level density $\delta g(kR)$
  as function of the wave-number variable $kR$
    was derived under these
    conditions for averaging parameter $\gamma$,
    or dimensional Gauss width $\Gamma$
  which are related to each other as described above.
According to the POT \cite{SM76,BB03,MY11},
the distance between major shells, $D_{\rm sh}$, is
determined by a
mean period of the  most 
short and degenerate POs, $\langle t^{}_{\rm PO}\rangle$
 (Refs.~\cite{SM76,BB03}):
\begin{equation}
\label{periodenp}
\mathcal{D}_{\rm sh} \cong 
\frac{2\pi \hbar}{\langle t^{}_{\rm PO}\rangle} 
\approx \frac{\lambda}{A^{1/3}}~.
\end{equation}
 The period, $A_{\rm sh}$, of the oscillating part, $\delta K(A)$, of the inverse
    level density parameter,
    $\delta K\propto -A\delta g/\tilde{g}^2 $ [see 
    Eq.~(\ref{a})
    for $a$ 
and Eq.~(\ref{goscsemnp}) for $\delta g(\lambda)$]
is defined approximately by the shell structure period
$\mathcal{D}_{\rm sh}$ of the 
level density $g$,
Eq.~(\ref{periodenp}),
as function of the 
 single-particle energy $\varepsilon$ near
  the Fermi surface,
$\varepsilon \approx \lambda$.
  Using the relationship between the particle number $A$ and the
    chemical potential $\lambda$ through
    Eq.~(\ref{conseq}), one obtains for the particle number period of the shell
    structure, $A_{\rm sh}$:
\begin{equation}
  \label{dA}
A_{\rm sh} \approx \mathcal{D}_{\rm sh}~\tilde{g} \sim 
A^{3/2}~,
\end{equation}
 where the TF estimate,
    $\tilde{g} \approx g^{}_{\rm TF}\sim A/\lambda$, was used.

Taking the factor in front of
$g^{}_{\rm PO}$ in Eq.~(\ref{dEPO0Fnp}) of the 
 shell correction energy
$\delta E_{\rm scl}$ 
    off the sum over the POs for the semiclassical energy-shell
correction \cite{SM76,SM77,MY11}, one arrives at
  \begin{equation}\label{dedgnp}
  \delta E_{\rm scl} 
  \approx
  \left(\frac{D_{\rm sh}}{2 \pi}\right)^2~ \delta g^{}_{\rm scl}(\lambda)~.
  \end{equation}
  Differentiating Eq.~(\ref{goscsemnp}) for $g_{\rm \tt{PO}}(\lambda)$
    over $\lambda$   in 
  Eq.~(\ref{dEPO0Fnp}), 
    one can keep only the dominating terms coming from 
    differentiation of the 
     cosine of the action phase argument,
$S/\hbar \sim A^{1/3}$. Finally, one finds the 
relationships which are useful in our
derivations (Sec.~\ref{MMA}): 
\begin{equation}
  \label{d2Edl2}
\frac{\partial^2\delta E_{\rm PO}}{\partial\lambda^2}
\approx -\delta g^{}_{\rm PO}~,\quad
\frac{\partial^2 g^{}_{\rm scl}}{\partial\lambda^2}\approx
\sum^{}_{\rm PO}\frac{\partial^2\delta g^{}_{\rm PO}}{\partial\lambda^2}
  \approx -\left(\frac{2\pi}{D_{\rm sh}}\right)^2 \delta g^{}_{\rm scl}(\lambda)~.
  \end{equation}


\begin{thebibliography}{99}

\bibitem{Be36} H.\ Bethe, Phys.\ Rev.\ {\bf 50} (1936) 332.  

\bibitem{Er60} T.\ Ericson, Adv.\ in Phys.\ {\bf 9} (1960) 425.

\bibitem{GC65} A.\ Gilbert, A.\ G.\ W.\ Cameron  Canadian J. of  Phys.
  {\bf 43} (1965) 1446. 

\bibitem{BM67} A.\ Bohr, B.\ R.\ Mottelson, {\it Nuclear structure}, vol. I
(Benjamin, New York, 1969).

\bibitem{St72} V.\ S.\ Stavinsky, Sov.\ J.\ Part.\ Nucl.\ {\bf 3} (1972) 417.

\bibitem{Bj74} S.\ Bj{\o}rnholm, A.\ Bohr, and B.\ R.\ Mottelson,
  {\it Role of symmetry of the nuclear shape in rotational contributions
    to nuclear level densities}: In
   Proc. of Symposium on the physics and chemistry of fission, Rochester, USA, 1973, vol. 1
    (Int. At. Energy Agency, Vienna, 1974), pp. 367-373.

\bibitem{Ig83} A.\ V.\ Ignatyuk, {\it Statistical properties of excited
  atomic nuclei} (Energoatomizadat, Moscow, 1983 (Russian)).

 \bibitem{So90} Yu.\ V.\ Sokolov, {\it Level density of atomic nuclei}
(Energoatomizadat, Moscow, 1990 (Russian)).
 
\bibitem{Sh92} S.\ Shlomo,
  Nucl.\ Phys.\ A{\bf 539} (1992) 17.

\bibitem{Ju98} A.\ R.\ Junghans, M.\ de Jong,
    H.-G.\ Clerc, A.\ V.\ Ignatyuk, G.\
 A.\ Kudyaev, K.-H.\ Schmidt,  Nucl.\ Phys.\ A {\bf 629}, 635 (1998). 

\bibitem{AB00} Y.\ Alhassid, G.\ F.\ Bertsch, S.\ Liu, H.\ Nakada, Phys.\
  Rev.\ C {\bf 84} (2000) 4313.

\bibitem{AB03} Y.\ Alhassid, G.F.\ Bertsch, 
L.\ Fang, Phys.\
  Rev.\ C {\bf 68} (2003) 044322. 

\bibitem{EB09} T.\ von Egidy, D.\ Bucurescu, 
  Phys.\ Rev.\ C {\bf 78} (2008) 051301(R);\\
   T.\ von Egidy, D.\ Bucurescu, Phys.\ Rev.\ C {\bf 80} (2009) 054310.

     \bibitem{Gr13}  S.\ M.\ Grimes,  Phys.\ Rev.\ C
       {\bf 88} (2013) 024613.

  \bibitem{AB15} Y.\ Alhassid, M.\ Bonett-Matiz, S.\ Liu, H.\ Nakada,
 Phys.\ Rev.\ C {\bf 92} (2015) 024307.     

  \bibitem{AB16} Y.\ Alhassid, G.\ F.\ Bertsch, C.\ N.\ Gilbreth, 
    H.\ Nakada, Phys.\
  Rev.\ C {\bf 93} (2016) 044320.

\bibitem{ZS16}  V.\ Zelevinsky, R.\ Senkov, Phys.\ Rev.\ C {\bf 93}
  (2016) 064304.  

\bibitem{KZ16} S.\ Karampagia, V.\ Zelevinsky,
    Phys.\ Rev.\ C {\bf 94}
 (2016) 014321.

\bibitem{KS18} V.M.\ Kolomietz, A.I.\ Sanzhur, S.\ Shlomo,
  Phys. Rev. C {\bf 97} (2018) 064302.
  
\bibitem{ZK18} V.\ Zelevinsky, S.\ Karampagia,
  Eur.\ Phys.\ J.\ Web\ Conf.\ {\bf 194} (2018) 01001.
  
\bibitem{ZH19} V.\ Zelevinsky, M.\ Horoi, 
  Prog.\ Part.\ Nucl.\ Phys.\ {\bf 105} (2019) 180.

\bibitem{Gr19} 
    S.\ M.\ Grimes, T.\ N.\ Massey, A.\ V.\ Voinov, Phys.\ Rev.\ C
  {\bf 99} (2019) 064331.

\bibitem{KZ20} S.\ Karampagia, V.\ Zelevinsky,  Int.\ J.\ Mod.\ Phys.\
  E {\bf 29} (2020) 2030005. 

\bibitem{KS20} V.\ M.\ Kolomietz, S.\ Shlomo, {\it Mean Field Theory}
  (World Scientific, 2020).

\bibitem{Ze96} V.\ Zelevinsky, B.\ A.\ Brown, N.\  Frazier,  M.\ Horoi, 
Phys.\ Rep.\ {\bf 276} (1996) 85.

\bibitem{Go11} J.\ M.\ G.\ Gomez, K.\ Kar, V.\ K.\ B.\ Kota,
  R.\ Molina, A.\ A.\ Relano,  J.\ Retamosa,
  Phys.\ Rep.\ {\bf 499} (2011) 103.

\bibitem{ML18}
  A.\ G.\ Magner, A.\ I.\ Levon, S.\ V.\ Radionov, 
  Eur.\ Phys.\ J.\  A {\bf 54} (2018) 214.

\bibitem{Le19a} A.\ I.\ Levon, D.\ Bucurescu, C.\ Costache,
  T.\ Faestermann, R.\ Hertenberger,
  A.\ Ionescu, R.\ Lica, A.\ G.\ Magner, C.\ Mihai, R.\ Mihai,
  C.\ R.\ Nita, S.\ Pascu,
  K.\ P.\ Shevchenko, A.\ A.\ Shevchuk, A.\ Turturica, H.-F.\ Wirth,
  Phys. Rev. C {\bf 102} (2020) 014308.

 \bibitem{Or97} W.\ E.\ Ormand, Phys.\ Rev.\ C {\bf 56} (1997)
   R1678.

   \bibitem{Ze16} F.\ Borgonovi, F.\ M.\ Israilev, L.\ F.\ Santos, 
  V.\ G.\ Zelevinsky, Phys.\ Rep.\ 
  {\bf 626} (2016) 1.

   \bibitem{Po65} C.\ E.\ Porter,
   {\it Statistical Theories of Spectra: Fluctuactions,}
   (Academic Press, 1965).

  
 \bibitem{Me04} M.\ L.\ Mehta, {\it Random Matrix Ensembles
   in Quantum Physics}, 3rd ed. (Elsevier,
  Amsterdam, 2004). 

\bibitem{KM79} 
V.\ M.\ Kolomietz, A.\ G.\ Magner, V.\ M.\ Strutinsky, 
Sov.\ J.\ Nucl.\ Phys.\ {\bf 29} (1979) 758.

 \bibitem{St67} V.\ M.\ Strutinsky,
    Nucl.\ Phys.\ A {\bf 95} (1967) 420;\\
     V.\ M.\ Strutinsky,
    Nucl.\ Phys. {\bf 122} (1968) 1.

\bibitem{BD72} 
  M.\ Brack, L.\ Damgaard, A.\ S.\ Jensen, A.\ C.\ Pauli,
  V.\ M.\  Strutinsky,  
  C.\ Y.\ Wong,  Rev.\ Mod.\ Phys.\ {\bf 44} (1972) 320.

 \bibitem{BK72} G.\ G.\ Bunatian, V.\ M.\ Kolomietz, V.\ M.\ Strutinsky,
  Nucl.\ Phys.\ A {\bf 188} (1972) 225.

 \bibitem{MI67} A.\ B.\ Migdal,
     {\it The Finite Fermi-System Theory and
Properties of Atomic Nuclei} (Intersience, New York,
1967; Ibid. Nauka, Moscow, 1983).

\bibitem{HS82} V.\ A.\ Khodel, E.\ E.\ Saperstein,  Phys.\ Rep.\ {\bf 5}
    (1982) 183.

  \bibitem{MS69} W.\ D.\ Myers, W.\ J.\ Swiatecki, Ann.\ Phys.\ (N.Y.)
    {\bf 55} (1969) 395;\\
   W.\ D.\ Myers, W.\ J.\ Swiatecki, Ann.\ Phys.\ (N.Y.) {\bf 84} (1974) 186.


\bibitem{BG85} M.\ Brack, C.\ Guet, H.-B.\ H{\aa}kansson,
  Phys.\ Rep.\ {\bf 123}
  (1985) 275.

\bibitem{BB03} 
M.\ Brack, R.\ K.\ Bhaduri, {\it Semiclassical Physics. 
Frontiers in Physics}, No. 96, 2nd ed. (Westview Press, Boulder, CO, 2003).  

\bibitem{BB72} R.\ Balian, C.\ Bloch, Ann.\ Phys.\ {\bf 69} (1972) 76.

\bibitem{SM76} 
V.\ M.\ Strutinsky, A.\ G.\ Magner,  Sov.\ J.\ Part.\ Nucl.\ {\bf 7} (1976)
138.

\bibitem{SM77} 
V.\ M.\ Strutinsky, A.\ G.\ Magner, S.\ R.\ Ofengenden, 
T.\ D{\o}ssing, Z.\ Phys.\ A {\bf 283} (1977) 269.

\bibitem{MY11} 
A.\ G.\ Magner, A.\ S.\ Yatsyshyn, K.\ Arita, M.\ Brack, Phys.\ At.\ Nucl.\
{\bf 74} (2011) 1445.

\bibitem{PRC} A.\ G.\ Magner, A.\ I.\ Sanzhur, S.\ N.\ Fedotkin,
  A.\ I.\ Levon, S.\ Shlomo, Phys.\ Rev.\  C, {\bf 104}  (2021)
  044319.

  \bibitem{IJMPE} A.\ G.\ Magner, A.\ I.\ Sanzhur, S.\ N.\ Fedotkin,
  A.\ I.\ Levon, S.\ Shlomo, Int.\ J.\ Mod.\ Phys.\ E {\bf 30} (2021)
  2150092.

   \bibitem{MSIS12}P.\ Moeller, A.J.\ Sierk, T.\ Ichikawa, H.\ Sagawa,
 {\it Atomic Data and Nuclear Data Tables, Atom.\ Data \& Nucl.\ Data Tables }
 {\bf 109-110}, 1-204 (2016); http://www.elsevier.com/locate/adt.   
    
 \bibitem{St58} V.\ M.\ Strutinsky, {\it On the nuclear level density in case of an energy gap}:
  In Proc. of Int. Conf. on Nucl. Phys. (Paris, 1958), pp. 617-622.

\bibitem{Ig72} A.\ V.\ Ignatyuk, Yu.\ V.\ Sokolov,
  Yadern. Fiz. {\bf 16} (1972) 217;
  Preprint FEI-327, (FEI, Obninsk, 1972), pp. 1-19.

\bibitem{DS73} W.\ Dilg, W.\ Shantl, M.\ Uhl,
  Nucl.\ Phys.\ A {\bf 217} (1973) 269.

\bibitem{ss1} B.\ K.\ Agrawal, S.\ Shlomo,  V.\ K.\ Au,
  Phys.\ Rev.\ C {\bf 72} (2005) 014310.

\bibitem{BM75} A.\ Bohr, B.\ R.\ Mottelson, {\it Nuclear Structure}, vol. 2
  (Benjamin, New York, 1975).

 
\bibitem{ENSDFdatabase}  
    National Nuclear Data Center On-Line Data Service
   for the ENSDF (Evaluated Nuclear Structure Data File) database,
   http://www.nndc.bnl.gov/ensdf.

  
\end{thebibliography}
\end{document}